\documentclass[onecolumn,showpacs]{revtex4}

\usepackage{graphicx}
\usepackage{dcolumn}
\usepackage{bm}

\input epsf

\begin{document}

\title{\large Emergent Universe and Phantom Tachyon Model}

\author{\bf Ujjal Debnath\footnote{ujjaldebnath@yahoo.com ,
ujjal@iucaa.ernet.in}}

\affiliation{Department of Mathematics, Bengal Engineering and\\
Science University, Shibpur, Howrah-711 103, India.}

\date{\today}

\begin{abstract}
In this work, I have considered that the universe is filled with
normal matter and phantom field (or tachyonic field). If the
universe is filled with scalar field,  Ellis et al have shown that
emergent scenario is possible only for $k=+1$ i.e. for closed
universe and here I have shown that the emergent scenario is
possible for closed universe if the universe contains normal
tachyonic field. But for phantom field (or tachyonic field), the
negative kinetic term can generate the emergent scenario for all
values of $k ~(=0,\pm 1)$. From recently developed statefinder
parameters, the behaviour of different stages of the evolution of
the emergent universe have been studied. The static Einstein
universe and the stability analysis have been briefly discussed
for both phantom and tachyon models.
\end{abstract}


\maketitle

Recently, Ellis and Maartens [1] have considered a cosmological
model where inflationary cosmologies exist in which the horizon
problem is solved before inflation begins, there is no big-bang
singularity, no exotic physics is involved and the quantum
gravity regime can even be avoided. The inflationary universe
emerges from a small static state that has within it the seeds
for the development of the microscopic universe and it is called
{\it Emergent Universe} scenerio (i.e., modern version and
extension of the Lemaitre-Eddington universe). They have shown
that the emergent scenario is possible only for $k=+1$ i.e., for
closed model. The universe has a finite initial size, with a
finite amount of inflation occurring over an infinite time in the
past and with inflation then coming to an end via reheating in the
standard way. There are several features for the emergent
universe: (i) the universe is almost static at the finite past
($t\rightarrow -\infty$) and isotropic, homogeneous at large
scales, (ii) It is ever existing and there is no timelike
singularity, (iii) the universe is always large enough so that
the classical description of space-time is adequate, (iv) the
universe may contain exotic matter so that the energy conditions
may be violated, (v) the universe is accelerating, etc. An
interesting example of this scenario is given by Ellis et al [2]
for a closed universe model with a minimally coupled scalar field
$\phi$ and special form of potential $V(\phi)$ with energy
density $\rho=\frac{1}{2}~\dot{\phi}^{2}+V(\phi)$ and pressure
$p=\frac{1}{2}~\dot{\phi}^{2}-V(\phi)$ and possibly some ordinary
matter with equation of state $p=w\rho$ where $-\frac{1}{3} \le
w\le 1$. There are several works on emergent universe scenarios
[3 - 5]. Mukherjee et al [6] have considered a general framework
for emergent universe model contains a fluid which has an EOS
$p=A\rho-B\sqrt{\rho}$ where $A$ and $B$ are constants. Also
Campo et al [7] have studied an emergent universe model in the
context of self-interacting Brans-Dicke theory. Very recently
Banerjee et al [8] have been obtained emergent universe
in brane world scenario.\\

In this work, I have considered that the universe is filled with
normal matter and phantom field [9] (or tachyonic field [10])
instead of normal scalar field. The Lagrangian of the tachyonic
field $\phi$ with potential $V(\phi)$ can be written as ${\cal
L}=-V(\phi)\sqrt{1-\epsilon\dot{\phi}^{2}}$ [10]. Phantom field
[9] has the property that it has negative kinetic term so that the
ratio between pressure and energy density is always less than
$-1$. Here $\epsilon=+ 1$ represents normal tachyon and
$\epsilon=-1$ represents phantom tachyon [10]. So my main
motivation is that if the universe is filled with phantom field
(or tachyonic field) instead of normal scalar field, the emergent
scenario is possible for flat, open and closed models.\\

For a FRW spacetime, the line element is

\begin{equation}
ds^{2} = - dt^{2} + a^{2}(t) \left[\frac{ dr^{2}}{1-k r^{2}} +
r^{2}(d\theta^{2}+\sin^{2}\theta d\phi^{2})\right]
\end{equation}

where $a(t)$ is the scale factor and $k ~(= 0, \pm 1)$ is the
curvature scalar. Now consider the Hubble parameter ($H$) and the
deceleration parameter ($q$) in terms of scale factor as

\begin{equation}
H=\frac{\dot{a}}{a}~~,~~q=-\frac{a\ddot{a}}{\dot{a}^{2}}=-1-\frac{\dot{H}}{H^{2}}
\end{equation}

We consider the universe contains normal matter and phantom field
(or tachyonic field). The Einstein equations for the space-time
given by equation (1) are
\begin{equation}
3H^{2}+\frac{3k}{a^{2}}=\rho_{m}+\rho_{\phi}
\end{equation}
and
\begin{equation}
2\dot{H}+3H^{2}+\frac{k}{a^{2}}= - (p_{m} + p_{\phi})
\end{equation}

where $\rho_{m}$ and $p_{m}$ are the energy density and pressure
of the normal matter connected by the equation of state
\begin{equation}
p_{m}=w\rho_{m}~,~~-1\le w\le 1
\end{equation}

and $\rho_{\phi}$ and $p_{\phi}$ are the energy density and
pressure due to the phantom field (or tachyonic field).
\\

Now consider there is no interaction between normal matter and
phantom field (or tachyonic field), so the normal matter and
phantom field (or tachyonic field) are separately conserved. The
energy conservation equations for normal matter and phantom field
(or tachyonic field) are

\begin{equation}
\dot{\rho}_{m} + 3 H( p_{m} + \rho_{m} ) = 0
\end{equation}
and
\begin{equation}
\dot{\rho}_{\phi} + 3 H( p_{\phi} + \rho_{\phi} ) = 0
\end{equation}

From equation (6) we have the expression for energy density of
matter as
\begin{equation}
\rho_{m} = \rho_{0} a^{-3 ( 1+ w)}
\end{equation}
where $ \rho_{0} $ is the integration constant. For emergent
universe, the scale factor can be chosen as [6]
\begin{equation}
a = a_{0}\left(\beta+e^{\alpha t}\right)^{n}
\end{equation}

where $a_{0},~\alpha,~\beta$ and $n$ are positive constants. So
the Hubble parameter and its derivatives are given by

\begin{equation}
H=\frac{n\alpha e^{\alpha t}}{\left(\beta+e^{\alpha t}\right)}~,~
\dot{H}=\frac{n\beta\alpha^{2}e^{\alpha t}}{\left(\beta+e^{\alpha
t}\right)^{2}}~,~\ddot{H}=\frac{n\beta\alpha^{3}e^{\alpha
t}(\beta-e^{\alpha t})}{\left(\beta+e^{\alpha t}\right)^{3}}
\end{equation}

Here $H$ and $\dot{H}$ are both positive, but $\ddot{H}$ changes
sign at $t=\frac{1}{\alpha}~\text{log}\beta$. Thus $H,~\dot{H}$
and $\ddot{H}$ all tend to zero as $t\rightarrow -\infty$. On the
other hand as $t\rightarrow \infty$ the solution gives
asymptotically a de Sitter universe.\\

\begin{figure}
\includegraphics[height=1.8in]{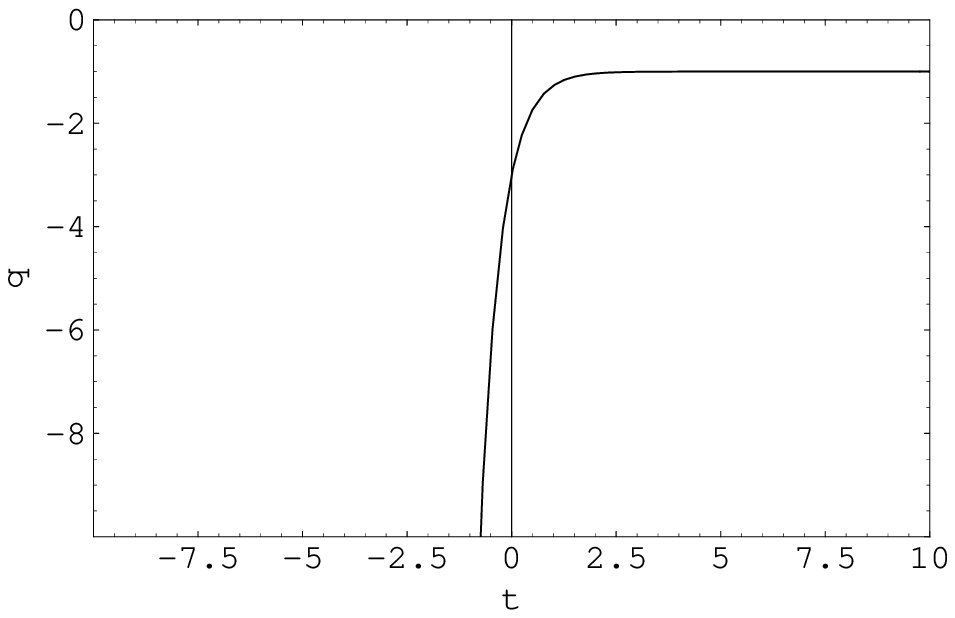}~~~~~
\includegraphics[height=1.8in]{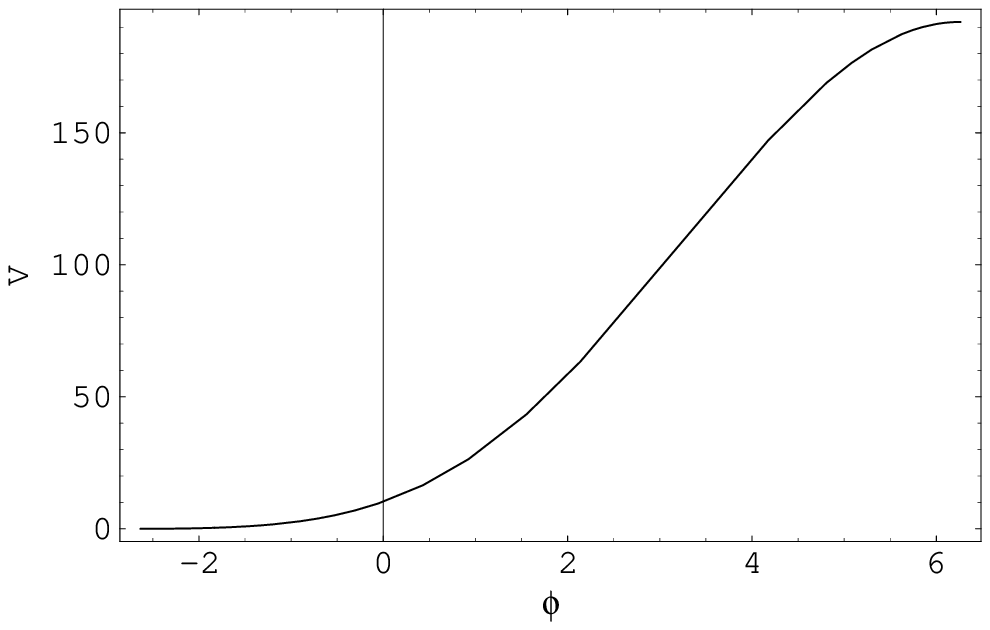}\\
\vspace{1mm} ~~~~~~~Fig.1~~~~~~~~~~~~~~~~~~~~~~~~~~~~~~~~~~~~~~~~~~~~~~~~~~~~~~~~~~~~~~~~~~~~~~~Fig.2\\

\vspace{6mm} Fig. 1 represents the variation of $q$ against $t$
for normalizing the constants $\alpha=2,~\beta=4$. Fig. 2 shows
the variation of $V$ against $\phi$ for normalizing the constants
$\alpha=2,~\beta=4,~w=1/3,~n=4,~\rho_{0}=3,~a_{0}=1$ in phantom
field.

\vspace{6mm}

\end{figure}

For the above choice of scale factor, the deceleration parameter
$q$ (see figure 1) can be simplified to the form

\begin{equation}
q=-1-\frac{\beta}{ne^{\alpha t}}
\end{equation}

$\bullet$ {\bf Phantom field:} The energy density $\rho_{\phi}$
pressure $p_{\phi}$ due to the phantom  field $\phi$ are given by

\begin{equation}
\rho_{\phi}=-\frac{1}{2}~\dot{\phi}^{2}+V(\phi)
\end{equation}
and
\begin{equation}
p_{\phi}=-\frac{1}{2}~\dot{\phi}^{2}-V(\phi)
\end{equation}

where $V(\phi)$ is the relevant potential for the phantom
field $\phi$. \\

From (3), (4), (12) and (13), we get

\begin{equation}
\dot{\phi}^{2}=2\dot{H}+(w+1)\rho_{m}-\frac{2k}{a^{2}}
\end{equation}
and
\begin{equation}
V(\phi)=\dot{H}+3H^{2}+\frac{1}{2}(w-1)\rho_{m}+\frac{2k}{a^{2}}
\end{equation}

From equations (5) and (10), it has been seen that $\dot{H}$ and
$(1+w)$ are always positive. Now the first and second terms of
equation (14) are always positive. So from eq. (14), it has been
seen that $\dot{\phi}^{2}>0$ for $k=0,-1$. But for $k=+1$,
$\dot{\phi}^{2}$ may or may not be positive. In this case,
$\dot{\phi}^{2}$ will be positive if
$2\dot{H}+(w+1)\rho_{m}>\frac{2k}{a^{2}}$ holds. So for phantom
model, the emergent scenario is possible for flat, open and
closed type of universes while for normal scalar field model the
emergent
scenario is possible only for closed universe [2].\\

From equations (3)-(13), one gets the expressions for $\phi$ and
$V$ as

\begin{equation}
\phi=\int \sqrt{(1+w)\rho_{0}a_{0}^{-3(1+w)} \left(\beta+e^{\alpha
t}\right)^{-3n(1+w)} +2n\beta\alpha^{2} e^{\alpha t}
\left(\beta+e^{\alpha t}\right)^{-2}  -2ka_{0}^{-2}
\left(\beta+e^{\alpha t}\right)^{-2n}   } ~ dt
\end{equation}
and
\begin{equation}
V=\frac{1}{2}(w-1)\rho_{0}a_{0}^{-3(1+w)} \left(\beta+e^{\alpha
t}\right)^{-3n(1+w)} +n\alpha^{2} e^{\alpha
t}\left(\beta+3ne^{\alpha t}\right) \left(\beta+e^{\alpha
t}\right)^{-2} +2ka_{0}^{-2} \left(\beta+e^{\alpha t}\right)^{-2n}
\end{equation}

Now it is very difficult to express the phantom field $\phi$ in
closed form, so potential function $V$ can not be expressed in
terms of $\phi$ explicitly. Now from the numerical
investigations, I have plotted $V$ against $\phi$ for some
particular values of arbitrary constants
($\alpha=2,~\beta=4,~w=1/3,~n=4,~\rho_{0}=3,~a_{0}=1$) in figure
2. From the figure, it is to be seen that $V$ is always increases
as $\phi$ increases from negative (at early universe) to positive
value (at late universe).\\\\

$\bullet$ {\bf Tachyonic field:} The energy density $\rho_{\phi}$
pressure $p_{\phi}$ due to the Tachyonic field $\phi$ have the
expressions

\begin{equation}
\rho_{\phi}=\frac{V(\phi)}{\sqrt{1-\epsilon{\dot{\phi}}^{2}}}
\end{equation}
and
\begin{equation}
p_{\phi}=-V(\phi) \sqrt{1-\epsilon{\dot{\phi}}^{2}}
\end{equation}

where $V(\phi)$ is the relevant potential for the tachyonic field
$\phi$. It is to be seen that $\frac{p_{\phi}}
{\rho_{\phi}}=-1+\epsilon\dot{\phi}^{2}$ $>-1$ or $<-1$ according
as normal tachyon ($\epsilon=+1$) or phantom tachyon
($\epsilon=-1$). From the tachyonic field, I have the expression
for $\dot{\phi}^{2}$ as
\begin{equation}
\dot{\phi}^{2}=\frac{\frac{2k}{a^{2}}-2\dot{H}-(1+w)\rho_{m}
}{\epsilon\rho_{\phi}}
\end{equation}

From equations (5) and (10), it has been seen that $\dot{H}$ and
$(1+w)$ are always positive. From eq. (20), it has been seen that
when $\epsilon=-1$, $\dot{\phi}^{2}$ is always positive for
$k=0,-1$. But for $k=+1$, $\dot{\phi}^{2}$ may or may not be
positive when $\epsilon=-1$. In this case, $\dot{\phi}^{2}$ will
be positive if $2\dot{H}+(w+1)\rho_{m}>\frac{2k}{a^{2}}$ holds.
Now when $\epsilon=+1$, $\dot{\phi}^{2}$ is always negative for
$k=0,-1$ but for $k=+1$, $\dot{\phi}^{2}$ will be positive if
$2\dot{H}+(w+1)\rho_{m}<\frac{2k}{a^{2}}$ holds. So from the above
discussion, it may be concluded that (i) if the universe contains
normal tachyonic field ($\epsilon=+1$), the emergent scenario is
possible only for closed universe and (ii) if the universe
contains phantom tachyonic field ($\epsilon=-1$), the emergent
scenario is
possible for flat, open and closed universes.\\

\begin{figure}
\includegraphics[height=1.8in]{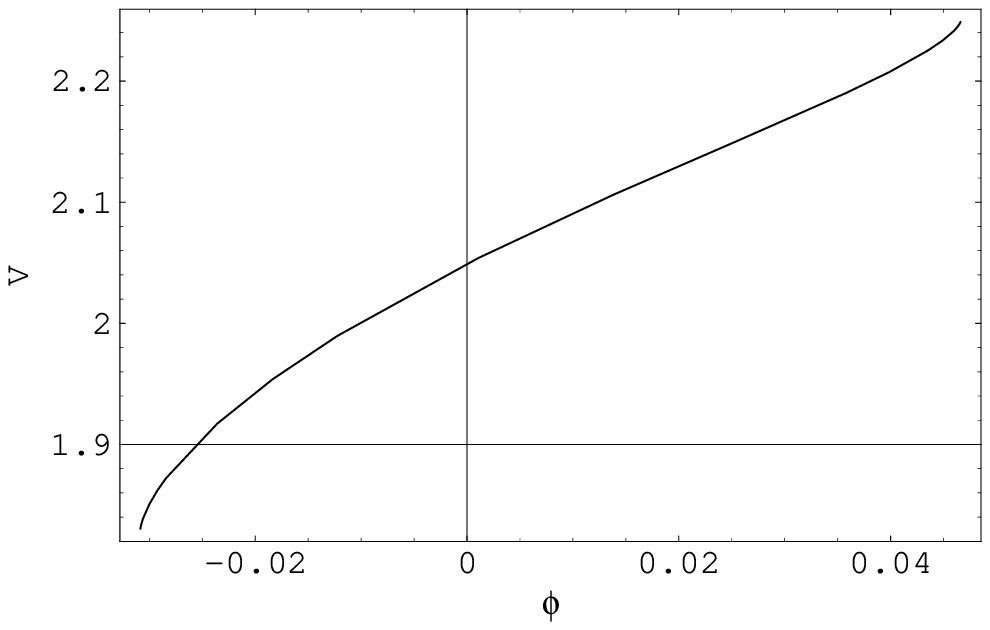}~~~~~
\includegraphics[height=1.8in]{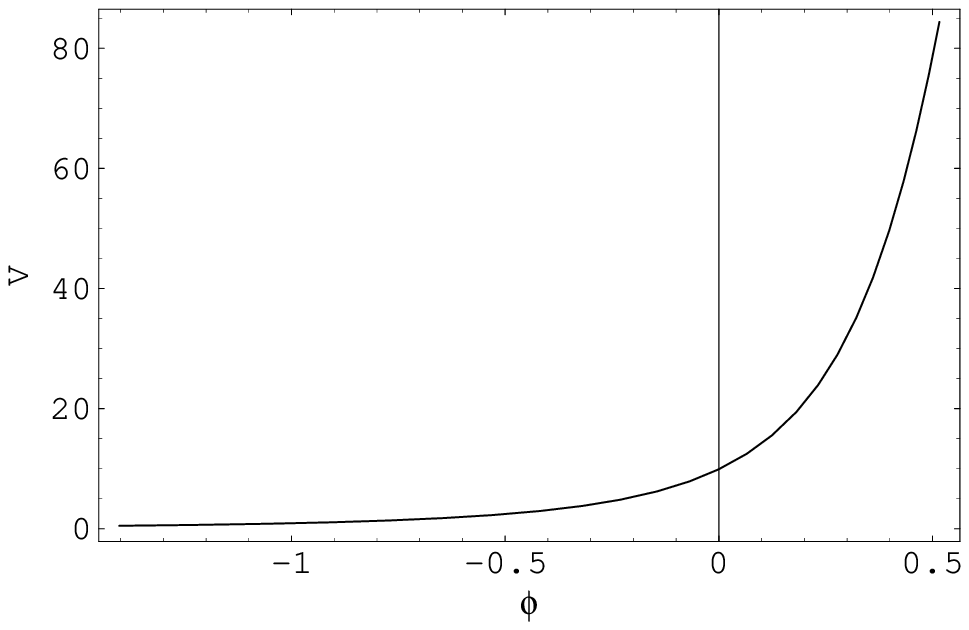}\\
\vspace{1mm} ~~~~~~~Fig.3~~~~~~~~~~~~~~~~~~~~~~~~~~~~~~~~~~~~~~~~~~~~~~~~~~~~~~~~~~~~~~~~~~~~~~~Fig.4\\

\vspace{6mm} Fig. 3 represents the variation of $V$ against $\phi$
for normalizing the constants $\alpha=1,~\beta=0.5,~w=1/3,~n=1$,
$~\rho_{0}=3,~a_{0}=1,~k=1$ in normal tachyon model
($\epsilon=+1$). Fig. 4 shows the variation of $V$ against $\phi$
for normalizing the constants
$\alpha=2,~\beta=4,~w=1/3,~n=4,~\rho_{0}=3,~a_{0}=1$ in phantom
tachyon model ($\epsilon=-1$).

\vspace{6mm}

\end{figure}

From equations (18)-(20), one gets the expressions for $\phi$ and
$V$ as

\begin{equation}
\phi=\int \sqrt{\frac{ka_{0}^{-2} \left(\beta+e^{\alpha
t}\right)^{-2n}-2n\beta\alpha^{2} e^{\alpha t}
\left(\beta+e^{\alpha t}\right)^{-2}-(1+w)\rho_{0}a_{0}^{-3(1+w)}
\left(\beta+e^{\alpha t}\right)^{-3n(1+w)}  }{\epsilon
\left\{3ka_{0}^{-2} \left(\beta+e^{\alpha t}\right)^{-2n}
+3n^{2}\alpha^{2} e^{2\alpha t} \left(\beta+e^{\alpha
t}\right)^{-2}-\rho_{0}a_{0}^{-3(1+w)} \left(\beta+e^{\alpha
t}\right)^{-3n(1+w)}\right\} }  } ~ dt
\end{equation}
and
\begin{eqnarray*}
V= \left[n\alpha^{2} e^{\alpha t}\left(2\beta+3ne^{\alpha
t}\right) \left(\beta+e^{\alpha t}\right)^{-2} +ka_{0}^{-2}
\left(\beta+e^{\alpha t}\right)^{-2n}-w\rho_{0}a_{0}^{-3(1+w)}
\left(\beta+e^{\alpha
t}\right)^{-3n(1+w)}\right]^{\frac{1}{2}}~\times
\end{eqnarray*}
\begin{equation}
\left[3ka_{0}^{-2} \left(\beta+e^{\alpha t}\right)^{-2n}
+3n^{2}\alpha^{2} e^{2\alpha t} \left(\beta+e^{\alpha
t}\right)^{-2}-\rho_{0}a_{0}^{-3(1+w)} \left(\beta+e^{\alpha
t}\right)^{-3n(1+w)}\right]^{\frac{1}{2}}
\end{equation}

Now it is very difficult to express the phantom field $\phi$ in
closed form, so potential function $V$ can not be expressed in
terms of $\phi$ explicitly. Now from the numerical
investigations, I have plotted $V$ against $\phi$ for some
particular values of arbitrary constants (i)
($\alpha=1,~\beta=0.5,~w=1/3,~n=1,~\rho_{0}=3,~a_{0}=1,~k=1$) in
figure 3 for normal tachyon model ($\epsilon=+1$) and
$\alpha=2,~\beta=4,~w=1/3,~n=4,~\rho_{0}=3,~a_{0}=1$ in figure 4
for phantom tachyon model ($\epsilon=-1$). From the figures, it is
to be seen that $V$ is always increases as $\phi$ increases from
negative to positive value.\\

In $2003$, V. Sahni etal [11] have introduced a pair of
parameters $\{r,s\}$, called statefinder parameters. In fact,
trajectories in the $\{r,s\}$ plane corresponding to different
cosmological models demonstrate qualitatively different
behaviour. These parameters can effectively differentiate between
different forms of dark energy and provide simple diagnostic
regarding whether a particular model fits into the basic
observational data. The above diagnostic pair has the following
form:
\begin{equation}
r = \frac{\dddot{a}}{aH^{3}} ~~\text{and}~~ s =
\frac{r-1}{3(q-\frac{1}{2})}
\end{equation}
For our model, the parameters $\{r,s\}$ can be explicitly written
in terms of $t$ as

\begin{equation}
r = 1+ \frac{\beta[\beta+(3n-1)e^{\alpha t}]}{n^{2}e^{2\alpha
t}}~~,~~ s=-\frac{2\beta[\beta+(3n-1)e^{\alpha
t}]}{3n[2\beta+3ne^{\alpha t}]e^{\alpha t}}
\end{equation}

OR, the relation between $r$ and $s$ has the form:

\begin{equation}
\{2\beta(r-1)-9s\}\{3(3n-2)s+2n\beta(r-1)\}-36ns^{2}(r-1)=0
\end{equation}

Figure 5 shows the variation of $s$ with the variation of $r$ for
for $\alpha=2,~\beta=4,~n=4$. For the emergent universe, $s$ is
always negative and $r\ge 1$. The curve shows that universe starts
from asymptotic Einstein static era ($r\rightarrow
\infty,~s\rightarrow -\infty$) and goes to $\Lambda$CDM model ($r=1,~s=0$).\\

\begin{figure}
\includegraphics[height=2.3in]{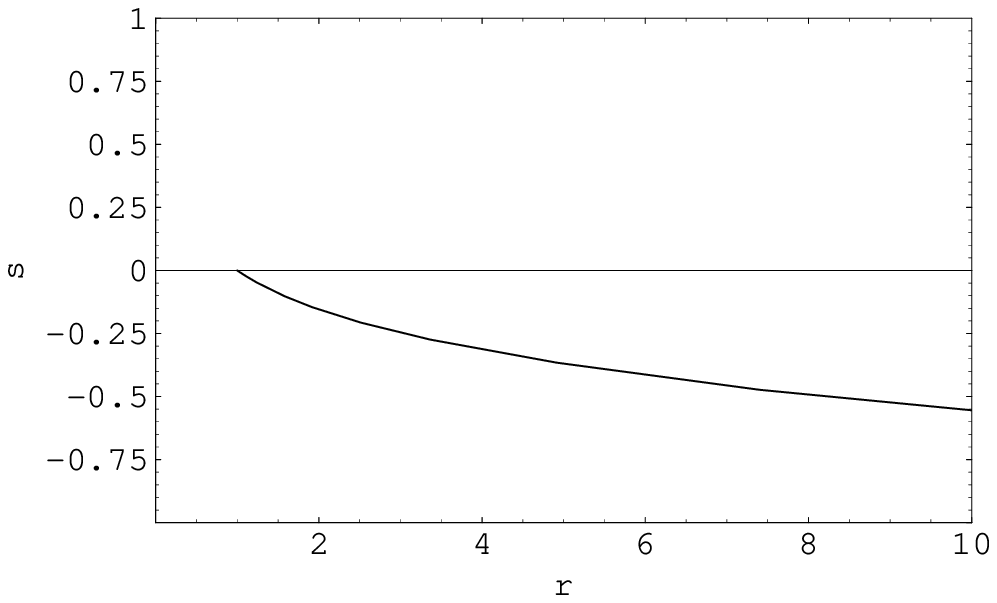}\\
\vspace{1mm} ~~~~~Fig.5\\

\vspace{6mm} Fig. 5 represents the variation of $s$ against $r$
for $\alpha=2,~\beta=4,~n=4$.

\vspace{6mm}

\end{figure}

The Einstein static universe is characterized by $k=+1,~a=$
constant. An initial Einstein static state of the universe arises
if the field $\phi$ starts out in an equilibrium position:
$V'(\phi_{0})=0$. Now consider, $V$ is trivial i.e., a flat
potential. Now general Einstein static model has
$\dot{\rho}_{m}=0$ and satisfies the following conditions
(using equations (3) and (4)):\\

$$
\frac{1}{2}(1+3w)\rho_{m0}-\dot{\phi}_{0}^{2}=V_{0}
$$
and
$$
(1+w)\rho_{m0}-\dot{\phi}_{0}^{2}=\frac{2}{a_{0}^{2}}
$$
for phantom model and
$$
\frac{1}{2}(1+3w)\rho_{m0}+\frac{(3\epsilon\dot{\phi}_{0}^{2}-2)V_{0}}{\sqrt{1-\epsilon\dot{\phi}_{0}^{2}}}=0
$$
and
$$
(1+w)\rho_{m0}+\frac{\epsilon\dot{\phi}_{0}^{2}V_{0}}{\sqrt{1-\epsilon\dot{\phi}_{0}^{2}}}=\frac{2}{a_{0}^{2}}
$$
for tachyon model.\\

Now it is easy to shown that if the field has no kinetic energy
($\dot{\phi}_{0}=0$), the equation of state is
$w_{\phi}=\frac{p_{\phi}}{\rho_{\phi}}=-1$ and if there is no
fluid ($\rho_{m0}=0$), the equation of state is
$w_{\phi}=-\frac{1}{3}$
for both phantom and tachyon models. Also it is easy to shown that
pure scalar field case is equivalent to the case of $w=1$.\\

Now consider the effect of inhomogeneous density perturbations on
the simple one-component fluid model. Following Gibbons [12] and
Barrow etal [13], it is easy to shown that the speed of sound
$c_{s}$ must satisfy the inequality
$c_{s}^{2}=\frac{dp_{m}}{d\rho_{m}}>\frac{1}{5}$ for stable case.
Harison [14] have discussed about stability for radiation filled
model and instability for dust filled model. Thus, Einstein static
universe with a fluid that satisfies the above inequality, is
neutrally stable against adiabatic density perturbations of the
fluid for inhomogeneous models with no scalar fields. Now consider
scalar field perturbation with non-flat potential $V(\phi)$ with
initial Einstein static state at $\phi=\phi_{0}$. Following Barrow
etal [13], it is easy to shown that the stability is not
significantly changed for scalar field perturbation in both
phantom
and tachyon models.\\

In this work, I have considered that the universe is filled with
normal matter and phantom field (or tachyonic field). Phantom
field has the property that it has negative kinetic term so that
the ratio between pressure and energy density is always less than
$-1$. It has been seen that the emergent scenario is possible for
flat, open and closed universes if the universe contains normal
phantom field or phantom tachyonic field. But if the universe
contains normal tachyonic field, the emergent scenario is possible
only for closed universe. In these cases, the field $\phi$ starts
from negative value at early stage and ends to positive value at
late stage. Figures 2-4 show that $V$ is increases as $\phi$
increases. For this emergent universe, $s$ is always negative and
decreases with $r$ increases. $\{r,s\}$ diagram (fig. 5) shows
that the evolution of emergent universe starts from asymptotic
Einstein static era ($r\rightarrow \infty,~s\rightarrow -\infty$)
and goes to $\Lambda$CDM model ($r=1,~s=0$).\\

{\bf Acknowledgement:}\\

The author is thankful to the authority of Institute of
Mathematical Sciences, Chennai, India for providing Associateship
Programme under which the work was carried out. Also the author
is thankful to UGC, Govt. of India for providing
research project grant (No. 32-157/2006(SR)).\\

{\bf References:}\\
\\
$[1]$ G. F. R. Ellis and R. Maartens, {\it Class. Quantum Grav.}
{\bf 21} 223 (2004).\\
$[2]$ G. F. R. Ellis, J. Murugan and C. G. Tsagas, {\it Class.
Quantum Grav.} {\bf 21} 233 (2004).\\
$[3]$ D. J. Mulryne, R. Tavakol, J. E. Lidsey and G. F. R. Ellis,
{\it Phys. Rev. D} {\bf 71} 123512 (2005).\\
$[4]$ J -H She, {\it JCAP} {\bf 0702} 021 (2007).\\
$[5]$ S. Mukherjee, B. C. Paul, S. D. Maharaj and A. Beesham,
{\it gr-qc}/0505103.\\
$[6]$ S. Mukherjee, B. C. Paul, N. K. Dadhich, S. D. Maharaj and
A. Beesham, {\it Class. Quantum Grav.} {\bf 23} 6927 (2006).\\
$[7]$ S. de Campo, R. Herrera and P. Labra$\tilde{\text{n}}$a, {\it JCAP} {\bf 0711} 030 (2007).\\
$[8]$ A. Banerjee, T. Bandyopadhyay and S. Chakraborty, {\it
Gravitation and Cosmology} {\bf 13} 290 (2007); {\it Gen. Rel. Grav.} {\bf 40} 1603 (2008).\\
$[9]$ B. Chang, H. Liu, L. Xu and C. Zhang, {\it Chin. Phys.
Lett.} {\bf 24} 2153 (2007).  ({\it
astro-ph}/0704.3768).\\
$[10]$ J. -g. Hao and X. -z. Li, {\it Phys. Rev. D} {\bf 68}
043510 (2003); {\it Phys. Rev. D} {\bf 68} 083514 (2003); S.
Nojiri and S. D. Odintsov, {\it Phys. Lett. B} {\bf 571} 1 (2003);
B. Gumjudpai, T. Naskar, M. Sami and S. Tsujikawa,
{\it JCAP} {\bf 0506} 007 (2005).\\
$[11]$ V. Sahni, T. D. Saini, A. A. Starobinsky and U.
Alam, {\it JETP Lett.} {\bf 77} 201 (2003).\\
$[12]$ G. W. Gibbons, {\it Nucl. Phys. B} {\bf 292} 784 (1987).\\
$[13]$ J. D. Barrow, G. F. R. Ellis, R. Maartens and C. G. Tsagas, {\it Class. Quantum Grav.} {\bf 20} L155 (2003).\\
$[14]$ E. R. Harison, {\it Rev. Mod. Phys.} {\bf 39} 862 (1967).\\

\end{document}